\documentstyle[prl,aps,epsfig,rotating]{revtex}

\begin{document}


\title{ Artificiality of multifractal phase transitions }

\author{
Martin Wolf$^{1}$, J\"urgen Schmiegel$^{1}$, and Martin Greiner$^{1,2}$
}

\address{$^1$Max-Planck-Institut f\"ur Physik komplexer Systeme, 
             N\"othnitzer Str.\ 38, D--01187 Dresden, Germany }
\address{$^2$Institut f\"ur Theoretische Physik, Technische Universit\"at,
             D--01062 Dresden, Germany }

\date{02.12.1999}

\maketitle

\begin{abstract}
A multifractal phase transition is associated to a nonanalyticity 
in the generalised dimensions. We show that its occurrence is an 
artifact of the asymptotic scaling behaviour of integral moments and 
that it is not observed in an analysis based on differential 
$n$-point correlation densities.
\end{abstract}

\vspace*{2cm}

\noindent
PACS: 02.50.Sk, 47.53.+n, 47.27.Eq, 05.45.+b \\
KEYWORDS:  multifractals, phase transition,
multiplicative branching processes, fully developed turbulence,
multivariate correlation densities. 

\vspace*{4cm}

\noindent
CORRESPONDING AUTHOR: \\
Martin Greiner \\
Max-Planck-Institut f\"ur Physik komplexer Systeme \\
N\"othnitzer Str.\ 38 \\ 
D--01187 Dresden, Germany  \\
tel.: 49-351-871-1218 \\
fax:   49-351-871-1199 \\
email: greiner @ mpipks-dresden.mpg.de

\newpage

Multifractal measures appear in a number of nonlinear physical
phenomena like turbulence \cite{FRI95,MEN91}, chaotic dynamical
systems \cite{SCH95} and high-energetic multiparticle dynamics
\cite{WOL96}, to name but a few. Due to the close analogy between
the multifractal formalism and statistical thermodynamics
\cite{HAL86,FEI86,FED88,MCC90,ARN95}, any nonanalyticity in the
generalised dimensions is interpreted as a multifractal phase
transition \cite{KAT87,SCH92,COL92}. This behaviour has been 
discussed in the context of the above mentioned phenomena
\cite{KAT87,BIA89,PES89,SCH94} and is also denoted as the occurrence
of strong intermittency. Note that conventionally generalised
dimensions are extracted from the asymptotic scaling behaviour 
of moments, where the latter represent integrals over the fundamental 
correlation densities. We will now demonstrate that a multifractal 
phase transition is an artefact of the integral moment analysis
and is not observed in an analysis based on differential correlation 
densities, where the true generalised dimensions of any order, 
characterising the underlying multiscale process, are revealed.

As the textbook example of a multifractal process we consider a
one-dimensional discrete random multiplicative cascade. It is
associated with a binary tree structure, obtained by hierarchically
partitioning the original interval of length $l_0=1$ into subintervals
of size $l_j=2^{-j}$. The density $\varepsilon^{(j+1)}_{\kappa}$,
associated with a $(j+1)$-generation interval characterised by the binary
index $\kappa = (k_1\cdots k_{j+1})$ with each $k$ taking on possible
values $0$ or $1$, is multiplicatively linked to the density at the 
larger scale by
\begin{equation}
\label{one}
  \varepsilon^{(j+1)}_{k_1\cdots k_{j}k_{j+1}}  
    =  q^{(j+1)}_{k_1\cdots k_{j}k_{j+1}} 
       \varepsilon^{(j)}_{k_1\cdots k_{j}}
       \quad .
\end{equation}
Independently for each branching, the left and right weights,
$q_L=q^{(j+1)}_{k_1\cdots k_{j}0}$ and $q_R=q^{(j+1)}_{k_1\cdots k_{j}1}$,
are drawn from a probabilistic splitting function $p(q_L,q_R)$
with support $0\leq q_L,q_R <\infty$. 
Without loss of generality we assume $p(q_L,q_R)=p(q_R,q_L)$  
and set $\langle q_L \rangle = 1$ as well as $\varepsilon^{(0)}=1$. 

Nature does not allow an infinitely long multifractal scaling range;
in fully developed turbulence, for example, it is restricted to
$\eta\ll l\ll L$, where $\eta$ and $L$ represent the Kolmogorov and the
integral length scales, respectively. Consequently, we restrict the
random multiplicative cascade to a finite number $J$ of cascade steps.
The complete statistical information of the ensemble of generated 
cascade fields is then contained in the multivariate characteristic 
function 
\begin{equation}
\label{two}
  Z[\lambda^{(J)}]
    =  \left\langle \exp \left(
       \sum_{k_1,\ldots,k_J=0}^{1}
       \lambda^{(J)}_{k_1\cdots k_J}
       \varepsilon^{(J)}_{k_1\cdots k_J}
       \right) \right\rangle
       \quad ,
\end{equation}
from which the $n$-point correlation densities are derived
\begin{equation}
\label{three}
  \rho^{[n]}_{\kappa_1\cdots\kappa_n}
    =  \left\langle
       \varepsilon^{(J)}_{\kappa_1} \cdots \varepsilon^{(J)}_{\kappa_n}
       \right\rangle
    =  \left.
       \frac{ \partial^n Z[\lambda^{(J)}] }
            { \partial\lambda^{(J)}_{\kappa_1} \cdots
              \partial\lambda^{(J)}_{\kappa_n} }
       \right|_{\lambda^{(J)}=0}
\end{equation}
by taking appropriate derivatives with respect to the conjugate field
variables $\lambda^{(J)}_{\kappa}$. The multivariate characteristic
function and the resulting $n$-point correlation densities have been
calculated analytically in Refs.\ \cite{GRE96,GIE97}; see also Refs.\
\cite{GRE98a,GRE98b}. 

For the extraction of generalised dimensions exponents so-called (box-) 
moments are considered, defined as
\begin{equation}
\label{four}
  M_n(J,j) 
    =  \frac{1}{2^j} \sum_{k_1,\ldots,k_j=0}^{1}
       \left\langle \left(
       \overline{\varepsilon}^{(J,j)}_{k_1\cdots k_j}
       \right)^n \right\rangle
       \quad ,
\end{equation}
where the backward density
\begin{equation}
\label{five}
  \overline{\varepsilon}^{(J,j)}_{k_1\cdots k_j}
    =  \frac{1}{2^{J-j}} \sum_{k_{j+1},\ldots,k_J=0}^{1}
       \varepsilon^{(J)}_{k_1\cdots k_J}
\end{equation}
has been resummed over the smallest scales from $J$ to $j$. 
With (\ref{five}) the (box-) 
moment (\ref{four}) can be understood as a (box-) integration over the
$n$-point correlation density (\ref{three}):
\begin{equation}
\label{six}
  \left\langle \left(
  \overline{\varepsilon}^{(J,j)}_{k_1\cdots k_j}
  \right)^n \right\rangle
    =  \frac{1}{2^{n(J-j)}}
       \sum_{k_{j+1}^{(1)},\ldots,k_J^{(1)}=0}^{1} \cdots
       \sum_{k_{j+1}^{(n)},\ldots,k_J^{(n)}=0}^{1}
       \left\langle
       \varepsilon^{(J)}_{k_1\cdots k_jk_{j+1}^{(1)}\cdots k_J^{(1)}}
       \cdots
       \varepsilon^{(J)}_{k_1\cdots k_jk_{j+1}^{(n)}\cdots k_J^{(n)}}
       \right\rangle
       \quad ,
\end{equation}
yielding the explicit expressions:
\begin{eqnarray}
\label{seven}
  M_1(J,j)
    &=&  1
         \quad , \nonumber \\
  M_2(J,j)
    &=&  \langle q^2 \rangle^j
         \left[
         \frac{ \langle q_Lq_R \rangle }{ 2 - \langle q^2 \rangle }
         + \frac{ 2 - \langle q^2 \rangle - \langle q_Lq_R \rangle }
                { 2 - \langle q^2 \rangle }
         \left( \frac{ \langle q^2 \rangle }{2} \right)^{J-j}
         \right]
         \quad ,
\end{eqnarray}
and, for arbitrary order,
\begin{equation}
\label{eight}
  M_n(J,j)
    =  \langle q^n \rangle^j
       \sum_{\{p\}} a_{\{p\}}^{(n)} \prod_{\Sigma n_i=n}
       \left( \frac{ \langle q^{n_i} \rangle }{2^{n_i-1}} \right)^{J-j} 
       \quad ,
\end{equation}
where $\{p\}$ stands for all possible partitions of $\sum_i n_i=n$ with 
$n_i \in \{1,2,\ldots,n\}$; the coefficients $a_{\{p\}}^{(n)}$ are simple
scale-independent functionals of the splitting-function moments
$\langle q_L^{m_1} q_R^{m_2} \rangle$ with $0\leq m_1+m_2 \leq n$.
--
While we will present a full technical understanding of the structure 
reflected in the expression (\ref{eight}) at a later stage of this 
presentation, some intuitive understanding can already be derived: 
the backward density (\ref{five}) can be rewritten as 
\begin{eqnarray}
\label{nine}
  \overline{\varepsilon}^{(J,j)}_{k_1\cdots k_j}
    &=&  q^{(1)}_{k_1} \cdots q^{(j)}_{k_1\cdots k_j}
         \frac{1}{2^{J-j}} \sum_{k_{j+1},\ldots,k_J=0}^{1}
         \varepsilon^{(J-j)}_{k_{j+1}\cdots k_J}
         \nonumber \\
    &=&  \varepsilon^{(j)}_{k_1\cdots k_j}
         \left( 1 + \Delta^{(J-j)}_{k_1\cdots k_j} \right)
         \quad ,
\end{eqnarray}
where $(1 + \Delta^{(J-j)})$ represents the resummed density of the
subcascade with length $J-j$ following the branching point 
$(k_1\cdots k_j)$. This resummed density need not be strictly equal to 
$1$ and in general will fluctuate around $1$. Hence, the backward
density (\ref{five}) need not be identical to the forward density
$\varepsilon^{(j)}_{k_1\cdots k_j}$. In view of (\ref{nine}), the first
factor $\langle q^n \rangle^j$ of the expression (\ref{eight}) then 
originates from $\langle (\varepsilon^{(j)}_{k_1\cdots k_j})^n \rangle$ 
while the remainder of the expression (\ref{eight}) is equal to 
$\langle (1+\Delta^{(J-j)})^n \rangle$.

For the very special case of a conservative cascade, where with 
$p(q_L,q_R)=p(q_L)\delta(q_L+q_R-2)$ the sum of the left and right 
weight is conserved at every branching, the two coefficients
$a_{\{p\}}^{(2)}$ of the second-order moment (\ref{seven}) become
$a_{\{1,1\}}
 = \langle q_Lq_R \rangle / (2 - \langle q^2 \rangle) 
 = 1$ 
and 
$a_{\{2,0\}}
 = (2 - \langle q^2 \rangle - \langle q_Lq_R \rangle)  
   / (2 - \langle q^2 \rangle)
 = 0$,
since $\langle q_Lq_R \rangle = \langle q(2-q) \rangle$. 
Similarly, all but one coefficients
$a_{\{p\}}^{(n)}$ of the expression (\ref{eight}) vanish and the moment of
order $n$ becomes exactly
$M_n(J,j) = \langle q^n \rangle^j$. Also in view of 
(\ref{nine}) this outcome is intuitively clear since due to the conservative 
nature of the splitting function the resummed density $1+\Delta^{(J-j)}$
is strictly equal to one. The moment $M_n(J,j)$ does 
not depend on the length $J$ of the cascade and exhibits perfect scaling.
The multifractal scaling exponents 
\begin{equation}
\label{ten}
  \tau(n)
    =  \frac{ \ln\langle q^n \rangle }{ \ln 2 }
\end{equation}
are deduced by setting $M_n(J,j) = (l_0/l_j)^{\tau(n)}$ 
and are related to the generalised dimensions $D_n$ by
$\tau(n) = (n-1)(1-D_n)$.

A factorised splitting function $p(q_L,q_R)=p(q_L)p(q_R)$ is a 
representative of non-conservative cascades, where the sum of the
left and right weight is only globally conserved 
($\langle q_L+q_R \rangle = 2$), but not locally ($q_L+q_R \neq 2$).
As a consequence, the mixed splitting-function moments 
$\langle q_L^{n_1}q_R^{n_2} \rangle 
 \neq  \langle q^{n_1}(2-q)^{n_2} \rangle$
do not show the anticorrelation typical of conservative cascades. 
For this case coefficients $a_{\{p\}}^{(n)}$ are generally nonzero and the
moments (\ref{eight}) do not show rigorous scaling. The resummed density
$1+\Delta^{(J-j)}$ of the subcascade with length $J-j$ now fluctuates 
around one and causes the deviations from rigorous scaling behaviour.

For nonconservative cascades two subclasses
of splitting functions have to be distinguished: 
in the case of so-called weak intermittency, the support of
$p(q_L,q_R)$ is restricted to $0\leq q_L,q_R < 2$, so that the respective
moments are restricted to $\langle q^n \rangle < 2^{n-1}$, where 
$n>1$ and where the extra
$1/2$ on the right hand side of this inequality comes from the requirement
$\langle q \rangle =1$. This implies that, in the limit of an infinitely 
long cascade, $J\rightarrow\infty$, only the true scaling term
$M_n(J,j) \sim \langle q^n \rangle^j$ survives in the expression 
(\ref{eight}), so that asymptotically for $j\ll J$ the same 
multifractal scaling exponents (\ref{ten})
are extracted as from the corresponding
conservative cascade. For the resummed density $1+\Delta^{(J-j)}$ of the
subcascade, this implies that in this asymptotic scaling range its 
probability distribution has converged to a scale-independent 
fixed point; see also Refs.\ \cite{BIA88,JOU99}.

The second subclass of nonconservative splitting functions exhibits 
the phenomenon that has become known as multifractal phase 
transition or strong intermittency: once the
support of $p(q_L,q_R)$ allows values $q_L$ and/or $q_R$ to
exceed $2$, a critical order $n_{\rm crit}$ exists, so that 
$\langle q^n \rangle / 2^{n-1} < 1$ for $n < n_{\rm crit}$ and 
$\langle q^n \rangle / 2^{n-1} > 1$ for $n > n_{\rm crit}$. 
Then, again in the limit of a very long, but finite cascade and given that 
$n>n_{\rm crit}$, the term corresponding to the partition
$\{n,0,\ldots,0\}$ dominates the moment (\ref{eight}) of order $n$,
which hence for $j\ll J$ scales as
\begin{equation}
\label{eleven}
  M_{n>n_{\rm crit}}(J,j\ll J)
    \approx  a^{(n)}_{\{n,0,\ldots,0\}}
               \left( \frac{ \langle q^n \rangle }{ 2^{n-1} } \right)^J
               2^{j(n-1)}
    \;\sim\; \left( \frac{ l_0 }{ l_j } \right)^{n-1}
               \quad . 
\end{equation}
For the multifractal scaling exponents this implies
\begin{equation}
\label{twelve}
  \tau(n) 
    =  \left\{
       \begin{array}{ll}
         \ln\langle q^n \rangle / \ln 2  &  \qquad (n<n_{\rm crit})  \\
         n-1  &  \qquad (n>n_{\rm crit}) \quad ,
       \end{array}
       \right.
\end{equation}
so that there is a discontinuity in the first derivative of $\tau(n)$
with respect to the moment-order $n$ at $n_{\rm crit}$. For 
$n<n_{\rm crit}$ the multifractal scaling exponents
$\tau(n) = \ln\langle \exp(n\ln q) \rangle / \ln 2$
may be interpreted as a free-energy-like function with moment
order $n$ as inverse temperature.
--
According to (\ref{eleven}), note that in the limit $J\rightarrow\infty$
of a non-physical, infinitely long cascade, moments with order larger than
$n_{\rm crit}$ would diverge. In view of (\ref{nine}) this implies that 
the probability distribution of the subcascade-resummed density 
$1+\Delta^{(J-j)}$ comes with an algebraic tail.

For a nonconservative random multiplicative cascade with a factorised
splitting function $p(q_L,q_R) = p(q_L)p(q_R)$, where, for example,  
\begin{equation}
\label{thirteen}
  p(q)
    =  \frac{ 1 }{ \sqrt{2\pi}\sigma q }
       \exp\left(
       -\frac{1}{2\sigma^2} \left( \ln q+\frac{\sigma^2}{2} \right)^2
       \right)
\end{equation}
is of log-normal type, the multifractal scaling exponents (\ref{twelve})
are found to be $\tau(n<n_{\rm crit})= \sigma^2 n(n-1) / (2\ln 2)$
below the critical order $n_{\rm crit} = 2\ln 2 / \sigma^2$,
which defines the multifractal phase transition at 
$\tau(n_{\rm crit}) = n_{\rm crit}-1$. In fully developed turbulence a
good qualitative description of observed multiplier distributions in 
the surrogate energy dissipation field \cite{JOU99} and an acceptable 
intermittency exponent $\tau(2)=0.25$ has been found for the parameter 
choice $\sigma=0.42$; for this value we find $n_{\rm crit} = 7.86$.
Note, that for the log-normal weight distribution the so-called 
Novikov rules \cite{NOV71} do not apply since weights may exceed a 
value of $2$ with nonzero probability.
--
The scale-dependence of the exact second-order moment expression 
(\ref{seven}) is illustrated in Fig.\ 1. As expected for $j\ll J$,
$M_2(J,j)$ scales as $\langle q^2 \rangle^j$ for 
$\sigma < \sigma_{\rm crit}$ and as $2^j$ for 
$\sigma > \sigma_{\rm crit}$, where $\sigma_{\rm crit} = \sqrt{\ln 2}$
corresponds to $n_{\rm crit}=2$. As $j\rightarrow J$ noticeable 
deviations from this scaling behaviour occur. Directly at 
$\sigma = \sigma_{\rm crit}$, where $\langle q^2 \rangle = 2$,
these finite size effects become so large
that it becomes difficult to extract a scaling exponent asymptotically.
Analogous findings hold for higher order moments $M_n(J,j)$.

In the presence of a multifractal phase transition the information
on the true multifractal scaling exponents (\ref{ten}) with 
$n > n_{\rm crit}$ appears to get lost. But this is not the case! 
Note in this respect that, according to (\ref{four})-(\ref{six}),
the moments $M_n(J,j)$ are box-integrals over the 
$n$-point correlation density (\ref{three}), the latter thus being more 
fundamental than the former. For demonstration, we pick the
two-point correlation density
\begin{equation}
\label{fourteen}
  \rho^{[2]}(d)
    =  \left\langle
       \varepsilon^{(J)}_{\kappa_1} \varepsilon^{(J)}_{\kappa_2}
       \right\rangle
    =  \left\langle q^2 \right\rangle^{J-d}
       \big(
       \langle q_Lq_R \rangle
       + \left( 1 - \langle q_Lq_R \rangle \right) \delta_{d,0}
       \big)
       \quad ,
\end{equation}
which is a function of the ultrametric distance $d=J-j$ between the 
two bins $\kappa_1=(k_1\cdots k_jk_{j+1}\cdots k_J)$ and
$\kappa_2=(k_1\cdots k_jk^\prime_{j+1}\cdots k^\prime_J)$, where
$k_{j+1}\neq k^\prime_{j+1}$.
It is depicted in Fig.\ 2 for a random multiplicative cascade with a
factorised splitting function of log-normal type and reveals perfect
scaling with the true multifractal scaling exponents. For the two-point
correlation density and, in general, for all $n$-point correlation 
densities no multifractal phase transition occurs. Then, why does a 
multifractal phase transition occur once based on integral moments and 
not on correlation densities?  

The answer to this question will be found in an inconspicuous property of
the correlation densities. For demonstration we discuss this 
again only for second order.
First, we realize that the second-order moment (\ref{four}) can be cast
in the form
\begin{equation}
\label{fivteen}
  M_2(J,j)
    =  \frac{ 1 }{ 2^{J-j} }
       \left(
       \rho^{[2]}(d=0)
       + \sum_{d=1}^{J-j} 2^{d-1} \rho^{[2]}(d)
       \right)
       \quad .
\end{equation}
Next, by adding and again subtracting something for $d=0$, the expression 
(\ref{fourteen}) is rewritten as
\begin{eqnarray}
\label{sixteen}
  \rho^{[2]}(d)
    &=&  \left\langle q^2 \right\rangle^{J-d}
         \left[ 
           \langle q_Lq_R \rangle
           + \left(
             \frac{ \langle q_Lq_R \rangle }{ 2 - \langle q^2 \rangle }
             - \langle q_Lq_R \rangle
           \right)  \delta_{d0}
         \right]
         + \frac{ 2-\langle q^2 \rangle -\langle q_Lq_R \rangle }
                { 2-\langle q^2 \rangle }
         \langle q^2 \rangle^J  \delta_{d0}
         \nonumber \\
    &\equiv&  
         \tilde{\rho}^{[2]}(d) 
         + \frac{ 2-\langle q^2 \rangle -\langle q_Lq_R \rangle }
                { 2-\langle q^2 \rangle }
         \langle q^2 \rangle^J  \delta_{d0}
         \quad .
\end{eqnarray}
For $d\neq 0$ the modified two-point correlation density
$\tilde{\rho}^{[2]}(d)$ is identical to the original two-point 
correlation density $\rho^{[2]}(d)$. The difference comes for $d=0$, 
where 
\begin{equation}
\label{seventeen}
  \frac{ \tilde{\rho}^{[2]}(d=0) }{ \tilde{\rho}^{[2]}(d=1) }
    =  \frac{ \langle q^2 \rangle }{ 2-\langle q^2\rangle }
\end{equation}
contrary to 
$\rho^{[2]}(d=0) / \rho^{[2]}(d=1)
 = \langle q^2 \rangle / \langle q_Lq_R \rangle$.
If we were only to substitute the modified two-point correlation density
into (\ref{fivteen}), then this difference insures perfect
moment scaling with the true multifractal scaling exponents 
as we arrive at the first term of the expression (\ref{seven}).
Substitution of only the second term appearing in (\ref{sixteen}),
which is proportional to $\delta_{d0}$, produces the second term
in (\ref{seven}), which is proportional to 
$(\langle q^2 \rangle / 2)^J 2^j$ and reflects a trivial scaling with
exponent $\tau(2)|_{\rm trivial} = 2-1$. The appearance of the second
term in (\ref{seven}) and (\ref{sixteen}), respectively, is a consequence
of the missing anticorrelation between weights $q_L$ and $q_R$ in the
splitting function; it vanishes only for a conservative splitting function
$p(q_L,q_R)=p(q_L)\delta(q_L+q_R-2)$ because then 
$\langle q_Lq_R \rangle = 2-\langle q^2 \rangle$. 

It is worthwhile to elaborate in more detail on the
occurrence of a multifractal phase transition from the viewpoint of
Eq.\ (\ref{sixteen}). For demonstration we pick again a factorised
splitting function of log-normal type (\ref{thirteen}), where 
$\langle q^2 \rangle = \exp(\sigma^2)$ and $\langle q_Lq_R \rangle = 1$.
As $\sigma$ increases monotonically
from $0$ to $\sigma_{\rm crit}=\sqrt{\ln 2}$, the
ratio $\langle q^2 \rangle / (2-\langle q^2\rangle)$ appearing in
(\ref{seventeen}) increases from $1$ to $+\infty$; then, as $\sigma$
is further increased, it changes sign and increases from $-\infty$ at
$\sigma=\sigma_{\rm crit}^+$ monotonically to $-1$ as 
$\sigma\rightarrow\infty$. Consequently, as the ($d=0$) and ($d=1$) 
elements of the modified two-point correlation density contribute as 
the sum $\rho^{[2]}(d=0)+\rho^{[2]}(d=1)$ to the second-order moment
(\ref{fivteen}), they enhance each other for $\sigma < \sigma_{\rm crit}$, 
but more or less cancel each other for $\sigma > \sigma_{\rm crit}$. 
Hence, for the former case the modified two-point correlation density 
dominates the moment scaling whereas for the latter case the 
$\delta$-function-like correction in (\ref{sixteen}) becomes dominant.

We conclude: when generalised dimensions are determined via integral
moments, one is likely to encounter artificial multifractal phase
transitions, which are not a property of the underlying strongly 
intermittent nonconservative cascade process, but artifacts of 
small-scale resummation. The fundamental $n$-point correlation 
densities and, should the underlying process only be resolvable 
at an intermediate scale, so-called density correlators 
\begin{equation}
\label{eighteen}
  {\cal C}^{[m_1,\ldots,m_n]}_{\kappa_1\cdots\kappa_n}
    =  \frac{ \left\langle 
              \left( \overline{\varepsilon}^{(J,j)}_{\kappa_1} 
                    \right)^{m_1}
              \cdots
              \left( \overline{\varepsilon}^{(J,j)}_{\kappa_n} 
                    \right)^{m_n} 
              \right\rangle }
            { \left\langle 
              \left( \overline{\varepsilon}^{(J,j)}_{\kappa_1} 
                    \right)^{m_1}
              \right\rangle 
              \cdots
              \left\langle 
              \left( \overline{\varepsilon}^{(J,j)}_{\kappa_n} 
                    \right)^{m_n}
              \right\rangle }
       \quad ,
\end{equation}
avoid such contributions and are therefore a better choice in 
estimating generalised dimensions.

\acknowledgements
The authors thank Bruno Jouault, Peter Lipa and Hans Eggers 
for some fruitful discussions. This work was supported in part 
by the Deutsche Forschungsgemeinschaft.

\newpage

\newpage
\begin{figure}
\begin{centering}
\begin{turn}{-90}
\epsfig{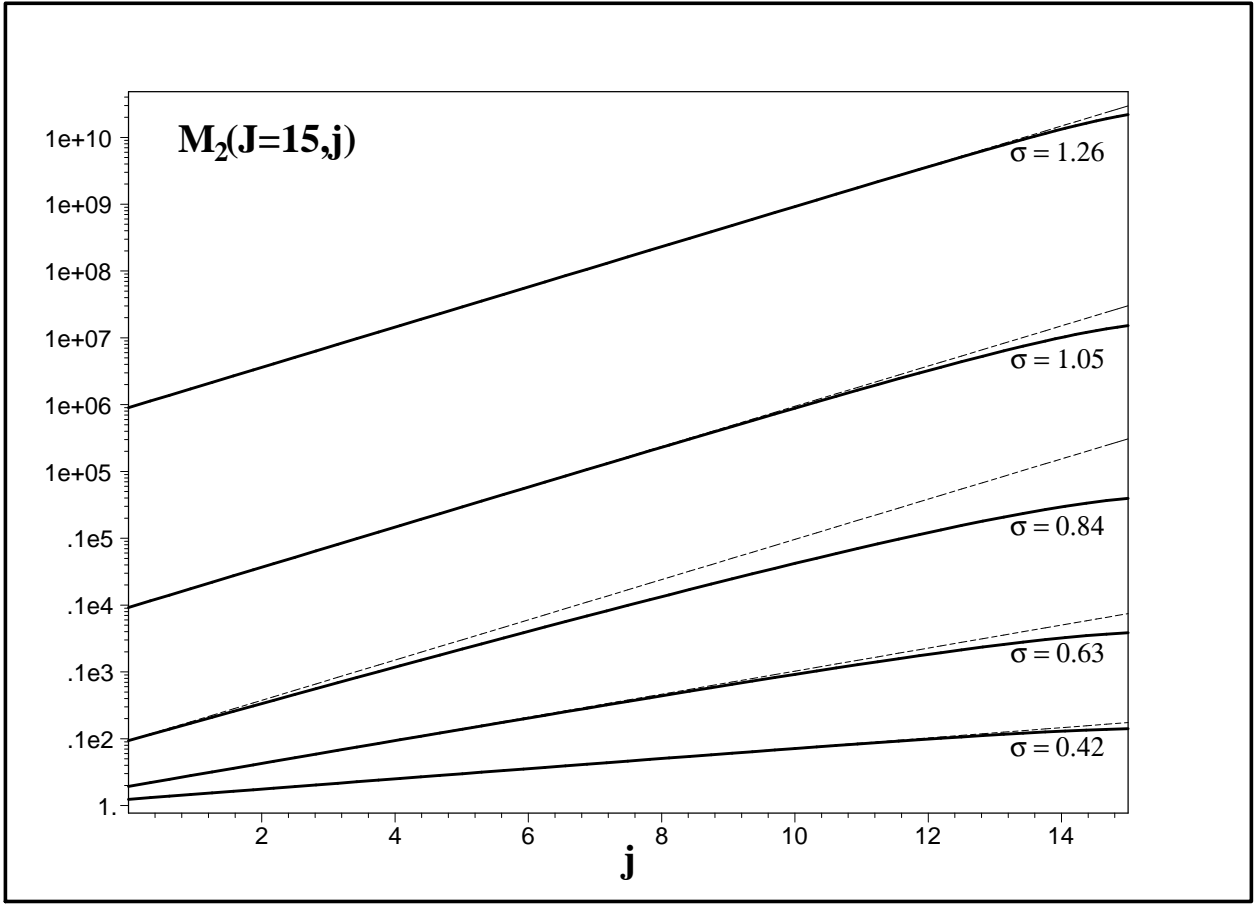}
\end{turn}\\[1cm]
\caption{
Scale-dependence of the second-order moment $M_2(J,j)$ resulting from
a finite discrete random multiplicative cascade with a factorised
log-normal splitting function. The dashed straight lines
represent the asymptotic scaling $M_2(J,j) \sim \langle q^2 \rangle^j$
and $M_2(J,j) \sim 2^j$ for 
$\sigma < \sigma_{\rm crit} = \sqrt{\ln 2} = 0.833$ and 
$\sigma > \sigma_{\rm crit}$, respectively.
} 
\end{centering}
\end{figure}

\newpage
\begin{figure}
\begin{centering}
\begin{turn}{-90}
\epsfig{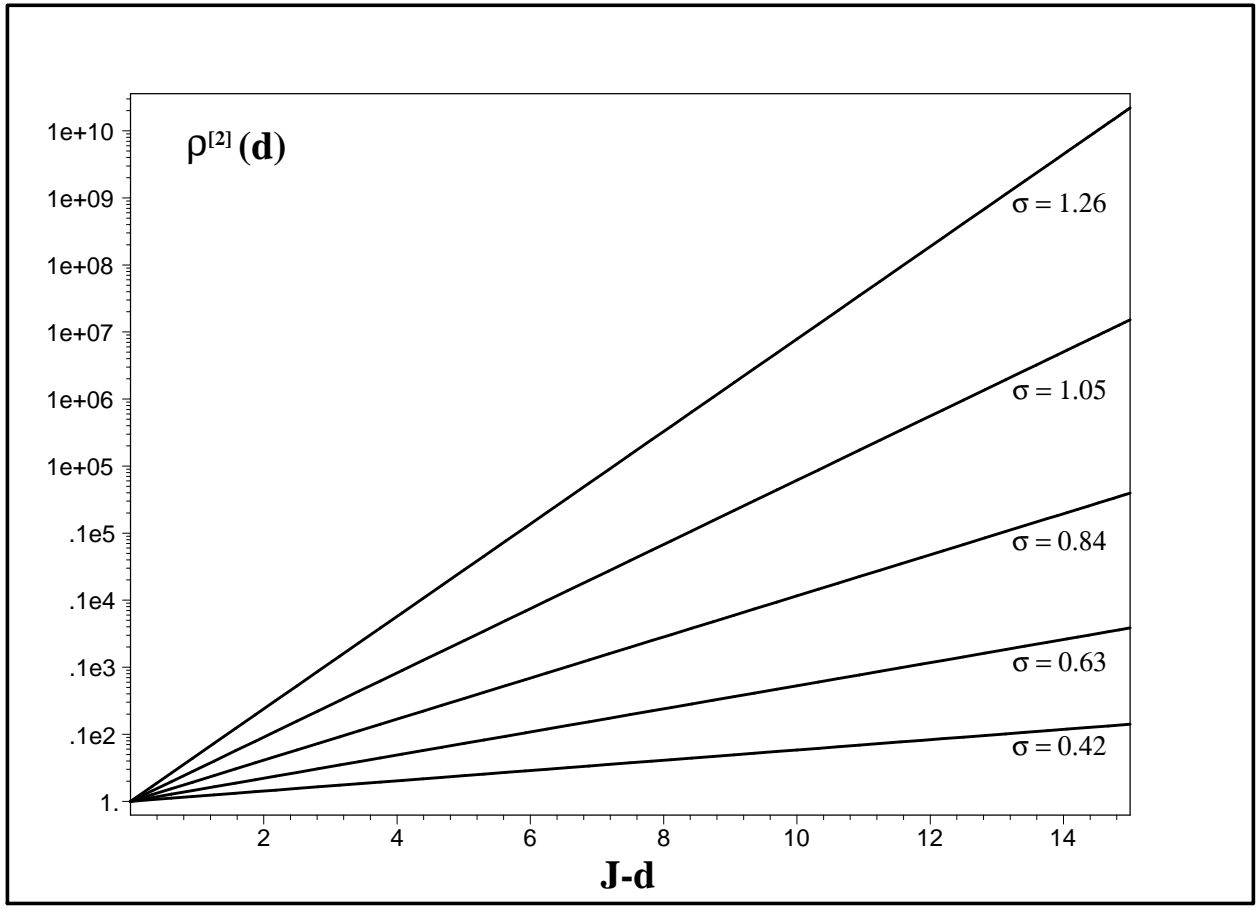}
\end{turn}\\[1cm]
\caption{
Dependence of the two-point correlation density $\rho^{[2]}(d)$, 
resulting from a finite discrete random multiplicative cascade 
with $J=15$ steps and a factorised log-normal splitting function, 
on the ultrametric distance. 
} 
\end{centering}
\end{figure}


\begin{thebibliography}{99}


\bibitem{FRI95}
         U.\ Frisch, 
         {\em Turbulence} 
         (Cambridge University Press, Cambridge, 1995).

\bibitem{MEN91}
         C.\ Meneveau and K.R.\ Sreenivasan,
         J.\ Fluid Mech.\ {\bf 224}, 429 (1991). 

\bibitem{SCH95}
         H.G.\ Schuster, 
         {\em Deterministic chaos: an introduction} 
         (VCH, Weinheim, 1995).

\bibitem{WOL96}
         E.A.\ De Wolf, I.M.\ Dremin and W.\ Kittel,
         Phys.\ Rep.\ {\bf 270}, 1 (1996).

\bibitem{HAL86}
         T.C.\ Halsey, M.H.\ Jensen, L.P.\ Kadanoff, I.\ Procaccia,
         and B.I.\ Shraiman,
         Phys.\ Rev.\ A {\bf 33}, 1141 (1986).

\bibitem{FEI86}
         M.J.\ Feigenbaum, M.H.\ Jensen, and I.\ Procaccia,
         Phys.\ Rev.\ Lett.\ {\bf 57}, 1503 (1986).

\bibitem{FED88}
         J.\ Feder, 
         {\em Fractals} 
         (Plenum Press, New York, 1988).

\bibitem{MCC90}
         J.L.\ McCauley,
         Phys.\ Rep.\ {\bf 189}, 225 (1990).

\bibitem{ARN95}
        A.\ Arneodo, E.\ Bacry, and J.F.\ Muzy,
        Physica A {\bf 213}, 232 (1995).

\bibitem{KAT87}
         D.\ Katzen, and I.\ Procaccia,
         Phys.\ Rev.\ Lett.\ {\bf 58}, 1169 (1987).

\bibitem{SCH92}
         D.\ Schertzer, and S.\ Lovejoy,
         Physica A {\bf 185}, 187 (1992).

\bibitem{COL92}
         P.\ Collet, and F.\ Koukiou,
         Commun.\ Math.\ Phys.\ {\bf 147}, 329 (1992).

\bibitem{BIA89}
         A.\ Bialas, and K.\ Zalewski,
         Phys.\ Lett.\ B {\bf 228}, 155 (1989).

\bibitem{PES89}
         R.\ Peschanski,
         Nucl.\ Phys.\ B {\bf 327}, 144 (1989).

\bibitem{SCH94}
         F.\ Schmitt, D.\ Schertzer, S.\ Lovejoy, and Y.\ Brunet,
         Nonlin.\ Proc.\ Geophys.\ {\bf 1}, 95 (1994).

\bibitem{GRE96}
        M.\ Greiner, J.\ Giesemann, P.\ Lipa, and P.\ Carruthers,
        Z.\ Phys.\ C {\bf 69}, 305 (1996).

\bibitem{GIE97}
        J.\ Giesemann, M.\ Greiner, and P.\ Lipa,
        Physica A {\bf 247}, 41 (1997).

\bibitem{GRE98a}
        M.\ Greiner, H.C.\ Eggers, and P.\ Lipa, 
        Phys.\ Rev.\ Lett.\ {\bf 80}, 5333 (1998).

\bibitem{GRE98b}
        M.\ Greiner, J.\ Schmiegel, F.\ Eickemeyer, P.\ Lipa, 
        and  H.C.\ Eggers, 
        Phys.\ Rev.\ E {\bf 58}, 554 (1998).

\bibitem{BIA88}
        A.\ Bialas, and R.\ Peschanski,
        Phys.\ Lett.\ B {\bf 207}, 59 (1988).

\bibitem{JOU99}
        B.\ Jouault, M.\ Greiner and P.\ Lipa,
        preprint chao-dyn/9812001,
        Physica D, in press.

\bibitem{NOV71}
        E.A.\ Novikov,
        Appl.\ Math.\ Mech.\ {\bf 35}, 231 (1971);
        Phys.\ Fluids A{\bf 2}, 814 (1990). 

\end{thebibliography}
\end{document}